\documentclass[twocolumn,showpacs,showkeys,reprint,amsmath,amssymb,aps]{revtex4-1}

\usepackage{graphicx}
\usepackage{dcolumn}
\usepackage{bm}
\usepackage{tabularx}
\usepackage{amsmath}

\begin{document}

\title{Heat transport study of the spin liquid candidate 1$T$-TaS$_2$}

\author{Y. J. Yu,$^{1,\dag}$ Y. Xu,$^{1,\dag}$ L. P. He,$^1$ M. Kratochvilova,$^{2,3}$ Y. Y. Huang,$^1$\\
J. M. Ni,$^1$ Lihai Wang,$^4$ Sang-Wook Cheong,$^4$ Je-Geun Park,$^{2,3}$ and S. Y. Li$^{1,5,*}$}

\affiliation
{$^1$State Key Laboratory of Surface Physics, Department of Physics, and Laboratory of Advanced Materials, Fudan University, Shanghai 200433, China\\
 $^2$Center for Correlated Electron Systems, Institute for Basic Science, Seoul 08826, Korea\\
 $^3$Department of Physics and Astronomy, Seoul National University, Seoul 08826, Korea\\
 $^4$Rutgers Center for Emergent Materials and Department of Physics and Astronomy, Rutgers University, Piscataway New Jersey 08854, USA\\
 $^5$Collaborative Innovation Center of Advanced Microstructures, Nanjing 210093, China
}

\date{\today}

\begin{abstract}
We present the ultra-low-temperature thermal conductivity measurements on single crystals of the prototypical charge-density-wave material 1$T$-TaS$_2$, which was recently argued to be a candidate for quantum spin liquid. Our experiments show that the residual linear term of thermal conductivity at zero field is essentially zero, within the experimental accuracy. Furthermore, the thermal conductivity is found to be insensitive to the magnetic field up to 9 T. These results clearly demonstrate the absence of itinerant magnetic excitations with fermionic statistics in bulk 1$T$-TaS$_2$ and, thus, put a strong constraint on the theories of the ground state of this material.
\end{abstract}

\maketitle

The quantum spin liquid (QSL), where strong quantum fluctuations obstruct long-range magnetic order even down to the absolute zero temperature, is one of the most elusive and exotic quantum state of matter \cite{Lee,Balents,YZhou}. In the QSLs, Mott physics plays a significant role in localizing electrons and forming $S = 1/2$ spins, as has been manifested in the study of high-temperature superconductors \cite{HTS review}. Experimentally, triangular-lattice organic compounds $\kappa$-(BEDT-TTF)$_2$Cu$_2$(CN)$_3$ \cite{Saito,Kanoda,k-salt kappa} and  EtMe$_3$Sb[Pd(dmit)$_2$]$_2$ \cite{Kato1,dmit kappa,Kato2}, together with kagome-lattice ZnCu$_3$(OH)$_6$Cl$_2$ \cite{Nocera,Lee1,Baines,Lee2,Lee3} and Cu$_3$Zn(OH)$_6$FBr \cite{Shi}, are typical examples of Mott-assisted QSL candidates. Richer physics, together with further complications, would be brought to the game, if this novel physics is to play on the stage of the charge-density-wave (CDW) state of a transition-metal dichalcogenide (TMD).

1$T$-TaS$_2$ is a layered material, and the only correlation-driven insulator discovered among TMDs \cite{Tutis}. As for the charge degree of freedom, 1$T$-TaS$_2$ features a number of peculiar CDW phases. Upon cooling, it turns into a metallic incommensurate CDW (ICCDW) phase below 550 K, a textured nearly commensurate CDW (NCCDW) phase below 350 K, and finally enters a commensurate CDW (CCDW) phase below 180 K \cite{Mott}. The low-temperature CCDW phase is characterized by a $\sqrt{13} \times \sqrt{13}$ structure described as star-of-David clusters \cite{Mott}. There is one unpaired electron per David-star due to energy gaps induced by the periodic lattice distortion \cite{Mott}. At the same time, the electron correlation effects set in and localize this electron, leading to a Mott insulating state with $S = 1/2$ spins arranged on an ideal triangular lattice \cite{Mott experiment1,Mott experiment2,Mott experiment3}. This is one of the few model spin configurations that may harbor the exotic QSL state, and exactly the one proposed by Anderson in his resonating-valence-bond model \cite{Anderson1,Anderson2,Baskaran}.

The possibility of the realization of a QSL in 1$T$-TaS$_2$ has been proposed recently in Ref. \cite{PALee proposal}. By analyzing the existing data of this material, it was argued that 1$T$-TaS$_2$ should be considered as a QSL, either a fully gapped $Z_2$ spin liquid or a Dirac spin liquid \cite{PALee proposal}. The muon spin relaxation ($\mu$SR) and nuclear quadrupole resonance (NQR) experiments have been performed on 1$T$-TaS$_2$ single crystals \cite{NQR}. No long-range magnetic order was detected from 210 K down to 70 mK by $\mu$SR. On the other hand, the NQR experiments reveal a gapless QSL-like behavior in part of the CCDW phase, from 200 K to $T_f$ = 55 K. Below $T_f$, a novel quantum phase with amorphous tiling of frozen singlets emerges out of the QSL \cite{NQR}. Meanwhile, another group performed polarized neutron diffraction and $\mu$SR measurements on 1$T$-TaS$_2$ \cite{Kratochvilova}. Their results indicate the presence of the short-ranged magnetic order below 50 K, and support the scenario that an orphan $S$ = 1/2 spin moment is localized at the center of the David-star \cite{Kratochvilova}.

To find out what is the true ground state of bulk 1$T$-TaS$_2$, it is essential to know the details of the low-lying elementary excitations. Ultra-low-temperature thermal conductivity measurement has proven to be a powerful technique in the study of low-lying excitations in QSL candidates \cite{k-salt kappa,dmit kappa,YMGO}. Taking the spin-1/2 triangular-lattice Heisenberg antiferromagnets as example, the thermal conductivity result implied a possibility of a tiny gap opening in $\kappa$-(BEDT-TTF)$_2$Cu$_2$(CN)$_3$ \cite{k-salt kappa}, while highly mobile gapless excitations with fermionic statistics exist in EtMe$_3$Sb[Pd(dmit)$_2$]$_2$ \cite{dmit kappa}. For the QSL candidate YbMgGaO$_4$, which has been studied extensively recently, no significant contribution of thermal conductivity from magnetic excitations was observed \cite{YMGO}.

In this Rapid Communication, we report the ultra-low-temperature thermal conductivity measurement on a high-quality 1$T$-TaS$_2$ single crystal down to 0.1 K. No significant contribution from magnetic excitations is detected at zero magnetic field. Furthermore, the thermal conductivity is found to be insensitive to magnetic fields up to 9 T. The absence of $\kappa_0/T$ at all fields unambiguously demonstrates that no fermionic magnetic excitations with itinerant character exist in 1$T$-TaS$_2$. We shall discuss the implications of our findings on the ground state of bulk 1$T$-TaS$_2$.

The high-quality 1$T$-TaS$_2$ single crystal was grown by the chemical vapor transport method \cite{Kratochvilova}. The x-ray diffraction (XRD) measurement was performed on the 1$T$-TaS$_2$ sample by using an x-ray diffractometer (D8 Advance, Bruker). The single crystal with a large natural surface was cut to a rectangular shape of 3.25 $\times$ 0.72 $\times$ 0.1 mm$^3$. The large natural surface (3.25 $\times$ 0.72 mm$^2$) was determined to be the (001) plane by XRD, as shown in Fig. 2(a). A standard four probe method was used for both resistivity and thermal conductivity measurements. Contacts were made directly on this natural surface with silver paint. The resistivity was measured in a $^4$He cryostat from 300 to 1.5 K. The thermal conductivity was measured in a dilution refrigerator, using a standard four-wire steady-state method with two RuO$_2$ chip thermometers, calibrated $in$ $situ$ against a reference RuO$_2$ thermometer. Magnetic fields were applied perpendicular to the large natural surface.

\begin{figure}
\includegraphics[clip,width=7cm]{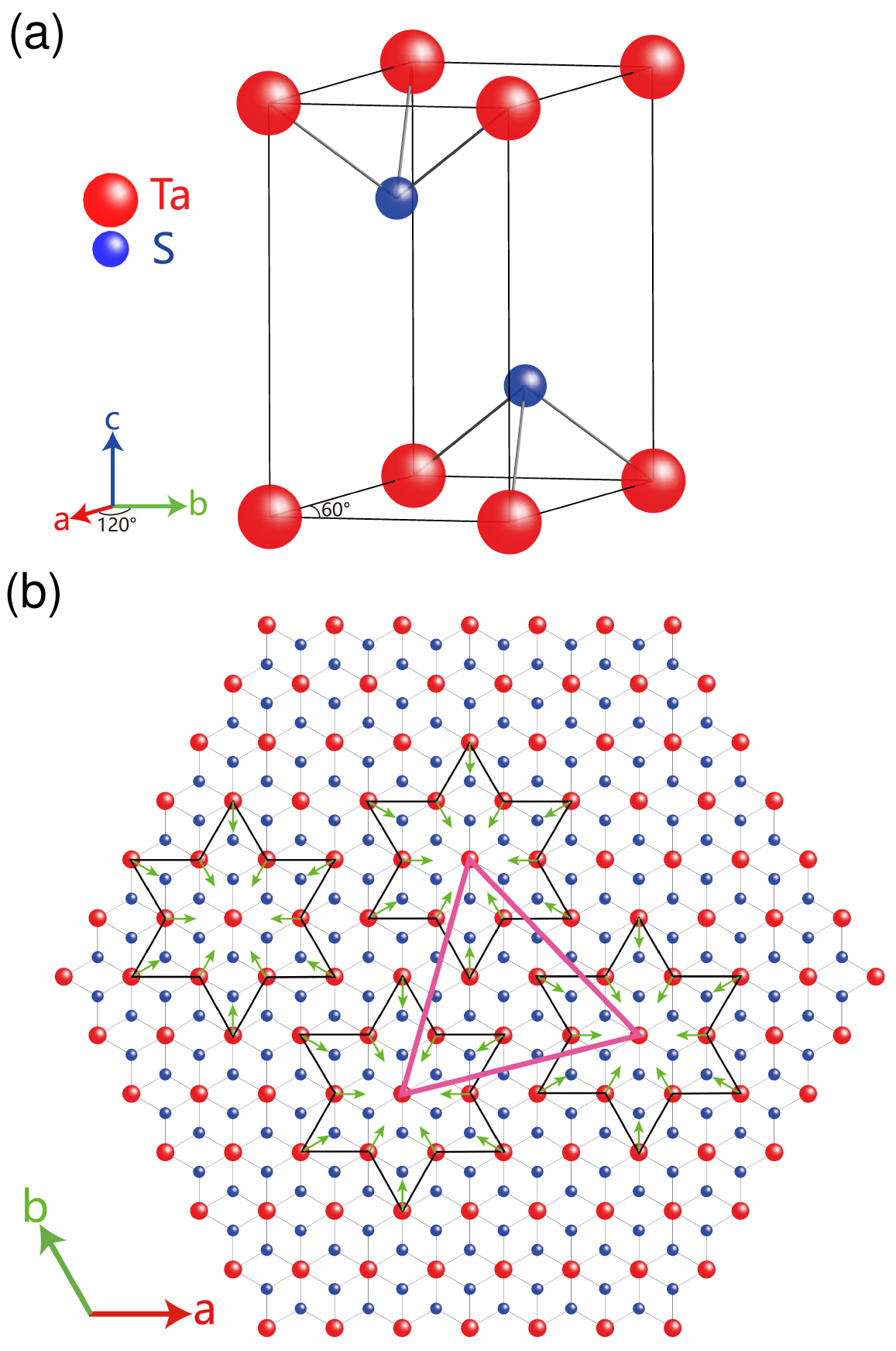}
\caption{(a) Trigonal $P$-$3m1$ crystal structure of 1$T$-TaS$_2$. Ta (S) atoms are displayed by red (blue) balls. (b) A schematic of the monolayer 1$T$-TaS$_2$ viewed along $c$ axis. Black line indicates the representative David-star clusters, where in-plane Ta atom displacements are marked by green arrows. These David-star clusters further form the triangular superlattice (magenta line).}
\end{figure}

As shown in Fig. 1(a), 1$T$-TaS$_2$ crystallizes in the CdI$_2$-type trigonal structure belonging to the $P$-$3m1$ space group \cite{Wiegers}. It has a layered structure, in which each atomic layer is composed of one Ta layer sandwiched between two S layers in an octahedral arrangement \cite{Clarke}. Within the CCDW phase, 13 Ta atoms form a fully interlocked David-star cluster, where 12 peripheral Ta atoms shrink towards the central Ta atom. Such a deformation leads to the formation of a $\sqrt{13} \times \sqrt{13}$ triangular superlattice \cite{Mott}, as illustrated in Fig. 1(b).

\begin{figure}
\includegraphics[clip,width=7.1cm]{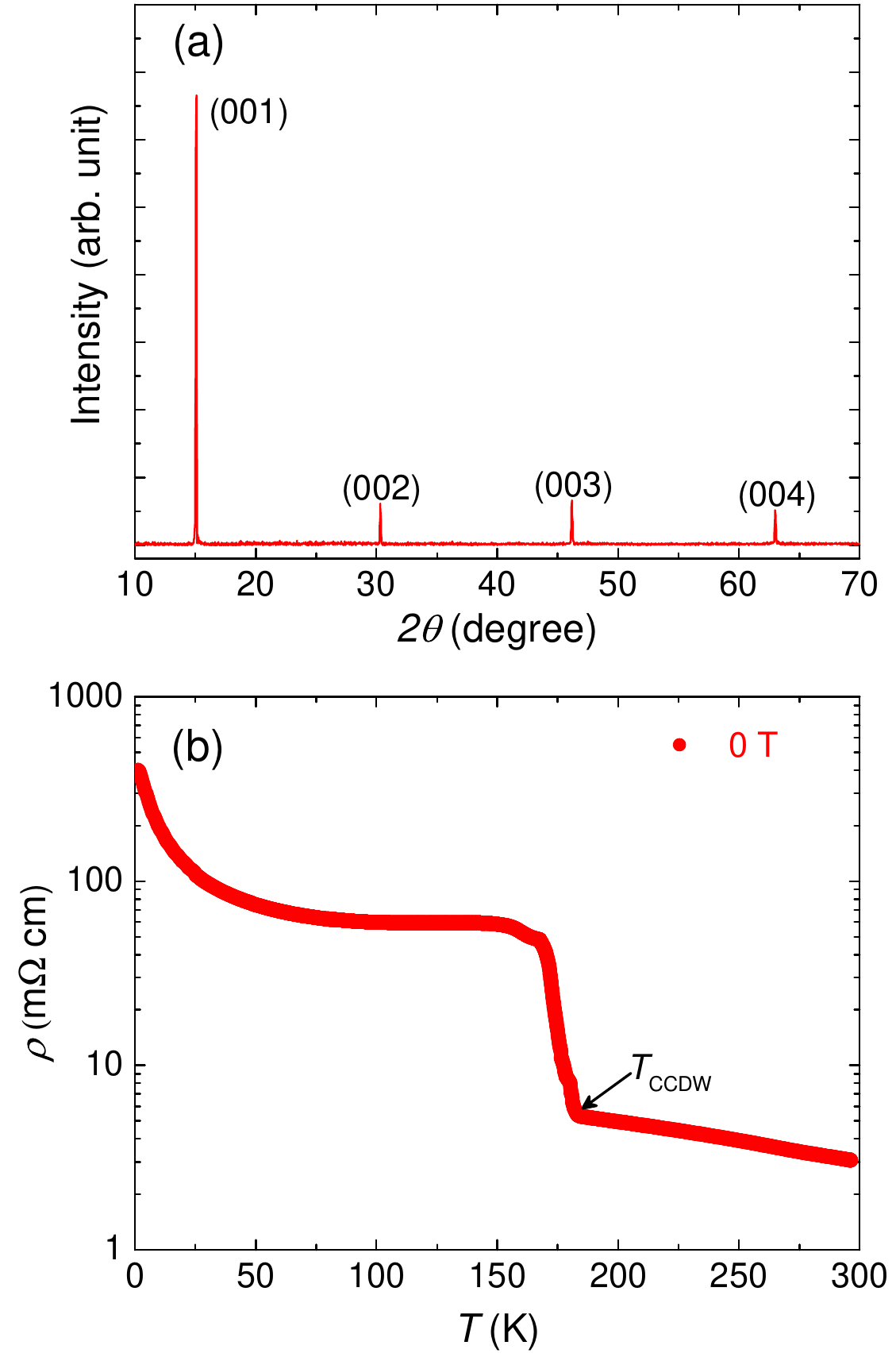}
\caption{(a) Room-temperature x-ray diffraction pattern from the large natural surface of the 1$T$-TaS$_2$ single crystal. Only (00$l$) Bragg peaks show up. (b) Temperature dependence of the resistivity (measured upon cooling) for the 1$T$-TaS$_2$ single crystal in zero magnetic field. The arrow indicates the onset of the transition from the NCCDW phase to the CCDW phase.}
\end{figure}

The temperature dependence of the resistivity $\rho(T)$ for the 1$T$-TaS$_2$ single crystal in zero magnetic field is plotted in Fig. 2(b). A sharp increase indicative of the occurrence of the transition from the NCCDW phase to the CCDW phase can be clearly seen around 180 K, below which the resistivity exhibits an insulating behavior. All of these features are consistent with previous resistivity measurements on 1$T$-TaS$_2$ \cite{Zhang,Iwasa}. The inverse resistivity ratio $\rho$(1.5 K)/$\rho$(295 K) is about 130, which is comparable with those measured previously \cite{Zhang,Iwasa}.

\begin{figure}
\includegraphics[clip,width=7.1cm]{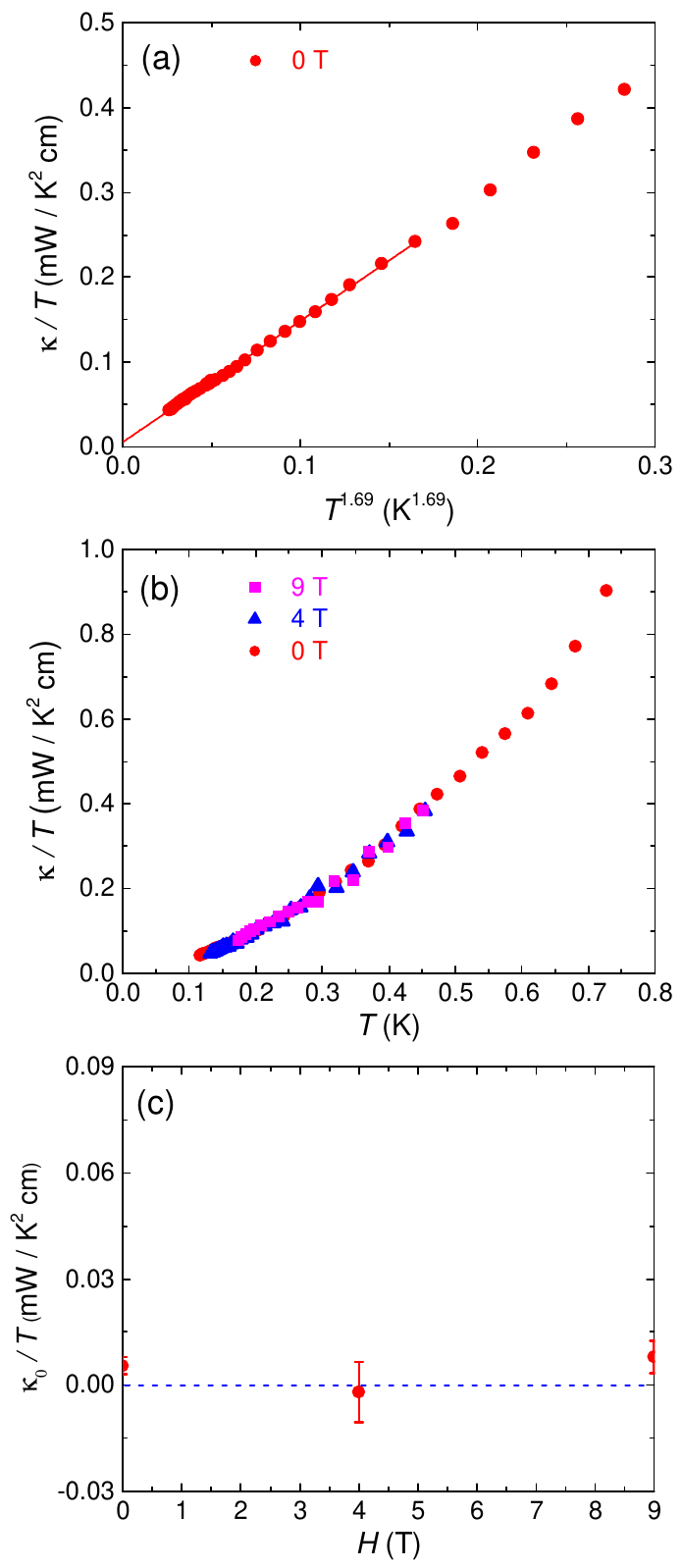}
\caption{(a) The in-plane thermal conductivity of the 1$T$-TaS$_2$ single crystal at $H$ = 0 T. The solid line represents the fit of the data to $\kappa/T$ = $a$ + $bT^{\alpha-1}$. This gives the residual linear term $\kappa_0/T$ = 0.005 $\pm$ 0.002 mW K$^{-2}$ cm$^{-1}$ and $\alpha$ = 2.69. (b) The in-plane thermal conductivity of 1$T$-TaS$_2$ at various magnetic fields ($H$ = 0, 4, and 9 T) applied along the $c$ axis. (c) Field dependence of the residual linear term $\kappa_0/T$. The three $\kappa_0/T$ values are negligible in our field range.}
\end{figure}

Figure 3(a) presents the in-plane thermal conductivity of the 1$T$-TaS$_2$ single crystal at $H$ = 0 T. In a solid, the contributions to thermal conductivity usually come from various quasiparticles, such as phonons, electrons, magnons, and spinons. For 1$T$-TaS$_2$, the thermal conductivity from electrons ($\kappa_{e}/T$) at 1.5 K is estimated to be 6.13 $\times$ 10$^{-5}$ mW K$^{-2}$ cm$^{-1}$ according to the Wiedemann-Franz law $\kappa_{e}/T$ = $L_0/\rho$(1.5 K), with the Lorenz number $L_0$ = 2.45 $\times$ 10$^{-8}$ W $\Omega$ K$^{-2}$ and $\rho$(1.5 K) = 399.8 m$\Omega$ cm. The electron contribution becomes smaller upon further cooling and is negligible at ultra-low temperature, due to the insulating behavior of the resistivity. Therefore, the thermal conductivity at very low temperature can be fitted by $\kappa/T$ = $a$ + $bT^{\alpha-1}$, in which the two terms $aT$ and $bT^{\alpha}$ represent the contributions from fermionic magnetic excitations (if they exist) and phonons, respectively \cite{Takagi,Taillefer}. Because of the specular reflections of phonons at the sample surfaces, the power $\alpha$ in the second term is typically between 2 and 3 \cite{Takagi,Taillefer}. The fitting of 0 T data below 0.35 K gives the residual linear term $\kappa_0/T$ $\equiv$ $a$ = 0.005 $\pm$ 0.002 mW K$^{-2}$ cm$^{-1}$ and $\alpha$ = 2.69. Considering our experimental error bar $\pm$ 5 $\mu$W K$^{-2}$ cm$^{-1}$, the $\kappa_0/T$ of 1$T$-TaS$_2$ at zero field is essentially zero. Note that EtMe$_3$Sb[Pd(dmit)$_2$]$_2$ has a value of $\kappa_0/T$ as big as 2 mW K$^{-2}$ cm$^{-1}$ \cite{dmit kappa}. The in-plane thermal conductivity of the 1$T$-TaS$_2$ single crystal in magnetic fields ($H$ = 0, 4, and 9 T) applied along the $c$ axis is plotted in Fig. 3(b), with the three curves almost overlapping on top of another. The same fitting process is performed, giving $\kappa_0/T$ = -0.002 $\pm$ 0.009 mW K$^{-2}$ cm$^{-1}$ and $\kappa_0/T$ = 0.008 $\pm$ 0.005 mW K$^{-2}$ cm$^{-1}$ for $H$ = 4 and 9 T, respectively. The three $\kappa_0/T$ values are plotted in Fig. 3(c). One can see that magnetic field barely has any effect on the thermal conductivity of 1$T$-TaS$_2$ up to 9 T.

Now we would like to discuss the implications of our thermal conductivity results on the proposal of 1$T$-TaS$_2$ being a QSL. Theoretically, all known QSLs can be classified in terms of a spectrum of gapless spinons (or their absence) and the nature of the emergent gauge fields to which they couple \cite{HYao}. Various kinds of exotic models have been proposed in the study of various QSL candidates \cite{YZhou}. A systematic analysis of whether these models can be applied to 1$T$-TaS$_2$ is beyond the scope of this work, and we only discuss the feasibility of these models in the light of our experimental data on the low-energy spin excitations. Generally, a finite residual linear term $\kappa_0/T$ represents the contribution to $\kappa$ from fermionic magnetic excitations in the zero temperature limit, i.e., the spectrum of the fermionic magnetic excitations is gapless. This might come from a spinon-Fermi surface or nodes in the momentum space. For 1$T$-TaS$_2$, the former one has been ruled out \cite{PALee proposal}, because of the tiny linear term $\gamma$ ($\sim$ 2 mJ mol$^{-1}$ K$^{-2}$) observed in specific heat \cite{Kratochvilova,Benda}. For the latter one, the most common case is a U(1) Dirac spin liquid \cite{U1Dirac,HTS review,Herbertsmithite review}. In such a state, nodal fermionic spinons at the Dirac points would still result in a finite $\kappa_0/T$, and the thermal conductivity would be enhanced by a magnetic field \cite{U1Dirac}. This is incompatible with our results that the $\kappa_0/T$ is negligible at all fields and the thermal conductivity is insensitive to magnetic field. It seems that any gapless QSL scenarios, whether gapless everywhere or only at nodes in the momentum space, are not consistent with our data. Note that there are also some exotic scenarios with nodal bosonic excitations \cite{YZhou,bosonic}. The contribution to the thermal conductivity from these nodal excitations exhibits a power-law temperature dependence ($\sim T^\delta$). However, unlike nodal fermionic excitations, for which the power-law exponent $\delta$ is 1, the $\delta$ value for nodal bosonic excitations is unknown in advance, so that it is hard to be separated from the phonon contribution.

However, there is another possibility that might reconcile our data with the gapless QSL scenarios. For the low-temperature phase ($T \leq T_f$ = 55 K) of 1$T$-TaS$_2$, the NQR shows a broad distribution of $1/T_1$ values with a stretched exponent $p < 1$ ($p \approx$ 0.5), implying a highly inhomogeneous magnetic phase at all Ta sites \cite{NQR}. We note that similar spectral broadening and stretched exponent behavior have been observed in another triangular-lattice QSL candidate $\kappa$-(BEDT-TTF)$_2$Cu$_2$(CN)$_3$ \cite{kappa salt NMR}. The thermal conductivity measurement on $\kappa$-(BEDT-TTF)$_2$Cu$_2$(CN)$_3$ also gives a negligible $\kappa_0/T$, which was argued to come possibly from the localization of the gapless spin excitations due to the inhomogeneity \cite{inhomogenity}. This might also be the case for 1$T$-TaS$_2$. The low-temperature phase ($T \leq T_f $) still exhibits a gapless behavior for the low energy fractional excitations, according to Ref. \cite{NQR}, but these gapless excitations can be localized so that they cannot conduct heat.

Next we turn to the gapped QSL scenarios. A fully gapped $Z_2$ spin liquid was suggested in Ref. \cite{PALee proposal}, which is a state with gapped spinons together with gapped visons. For $\kappa$-(BEDT-TTF)$_2$Cu$_2$(CN)$_3$, an alternative explanation of the negligible $\kappa_0/T$ is that the spin excitations are gapped \cite{k-salt kappa}. The total thermal conductivity of $\kappa$-(BEDT-TTF)$_2$Cu$_2$(CN)$_3$ is the sum of a phonon contribution term and a magnetic part with an exponential temperature dependence, indicating the existence of a gap ($\Delta \sim$ 0.46 K) in the spin excitation spectrum \cite{k-salt kappa}. A field-induced gap closing was also observed in $\kappa$-(BEDT-TTF)$_2$Cu$_2$(CN)$_3$ for magnetic fields higher than $\sim$ 4 T \cite{inhomogenity}. For 1$T$-TaS$_2$, however, we found that a similar fitting procedure does not work. Having said that, we caution that our data do not contradict the gapped QSL scenario. Indeed, if the magnitude of the gap is sufficiently large, the magnitude of the exponential term in the total thermal conductivity would be too small to be discerned at such low temperature, so that the total thermal conductivity is dominated by the phonon term. And the gap, if it exists, cannot be closed by a magnetic field up to 9 T. For example, an unusually large exchange interaction $J$ $\approx$ 0.13 eV ($\sim$ 1500 K) has been derived from the susceptibility data \cite{NQR}. The spin gap is estimated to be above 200 K in Ref. \cite{PALee proposal}. In fact, in most cases a large gap is more common than a small gap (compared to its $J$) as possibly in $\kappa$-(BEDT-TTF)$_2$Cu$_2$(CN)$_3$ \cite{k-salt kappa}.

Of course, our results do not entirely rule out the possibility that the ground state of bulk 1$T$-TaS$_2$ may not be a QSL. The low-temperature phase ($T \leq T_f $) proposed by the NQR experiment is a highly unusual one, featuring frozen singlets, pseudogap in the spinon density of states, and a high degree of local disorder \cite{NQR}. In Ref. \cite{Kratochvilova}, the neutron diffraction and $\mu$SR results provide evidence for the existence of a short-range-ordered state. To what extent do the behaviors of these states resemble those of a QSL is an open question and requires future scrutiny. As stated in Ref. \cite{PALee proposal}, it might be more interesting to look for a QSL ground state in ultra-thin crystals of 1$T$-TaS$_2$.

In summary, we have measured the thermal conductivity of a 1$T$-TaS$_2$ single crystal down to 0.1 K. No residual linear term of thermal conductivity was observed at zero field. The thermal conductivity is found to be insensitive to a magnetic field up to 9 T. These results provide evidence for the absence of itinerant magnetic excitations obeying fermionic statistics in 1$T$-TaS$_2$. Our results set strong constraints on the nature of its ground state and, thus, of its theoretical description.

This work is supported by the Ministry of Science and Technology of China (Grant No: 2015CB921401 and 2016YFA0300503), the Natural Science Foundation of China, the NSAF (Grant No: U1630248), the Program for Professor of Special Appointment (Eastern Scholar) at Shanghai Institutions of Higher Learning, and STCSM of China (No. 15XD1500200). The work in Korea was supported by the Institute for Basic Science (IBS) in Korea (IBS-R009-G1). The work at Rutgers University was supported by the NSF under Grant No. NSF-DMREF-1233349.\\

\noindent $^\dag$ Y. J. Yu and Y. Xu contributed equally to this work.
\noindent $^*$ E-mail: shiyan$\_$li@fudan.edu.cn.

\end{document}